\def\papertitle{Fast Differentiable Modal Simulation of Non-linear Strings, Membranes, and Plates}
\def\paperauthorA{Rodrigo Diaz}
\def\paperauthorB{Mark Sandler}

\documentclass[twoside,a4paper]{article}
\usepackage{etoolbox}
\usepackage{dafx_25}
\usepackage{amsmath,amssymb,amsfonts,amsthm}
\usepackage{euscript}
\usepackage[T1]{fontenc}
\usepackage[utf8]{inputenc}
\usepackage{ifpdf}
\usepackage[english]{babel}
\usepackage{caption}
\usepackage{subfig} 
\usepackage{color}
\usepackage{listings}

\input glyphtounicode
\ninept

\newcounter{numauth}
\newcounter{listcnt}
\newcommand\authcnt[1]{\ifdefined#1 \stepcounter{numauth} \fi}

\newcommand\addauth[1]{
\ifdefined#1 
\stepcounter{listcnt}
\ifnum \value{listcnt}<\value{numauth}
\appto\authorslist{, #1}
\else
\appto\authorslist{~and~#1}
\fi
\fi}
\authcnt{\paperauthorB}
\authcnt{\paperauthorC}
\authcnt{\paperauthorD}
\authcnt{\paperauthorE}
\authcnt{\paperauthorF}
\authcnt{\paperauthorG}
\authcnt{\paperauthorH}
\authcnt{\paperauthorI}
\authcnt{\paperauthorJ}
\def\authorslist{\paperauthorA}
\addauth{\paperauthorB}
\addauth{\paperauthorC}
\addauth{\paperauthorD}
\addauth{\paperauthorE}
\addauth{\paperauthorF}
\addauth{\paperauthorG}
\addauth{\paperauthorH}
\addauth{\paperauthorI}
\addauth{\paperauthorJ}

\usepackage{times}

\newif\ifpdf
\ifx\pdfoutput\relax
\else
   \ifcase\pdfoutput
      \pdffalse
   \else
      \pdftrue
   \fi
\fi

\ifpdf 
  \usepackage[pdftex,
    pdftitle={\papertitle},
    pdfauthor={\authorslist},
    pdfsubject={Proceedings of the 28th International Conference on Digital Audio Effects (DAFx25)},
    colorlinks=false, 
    bookmarksnumbered, 
    pdfstartview=XYZ 
  ]{hyperref}
  \usepackage[pdftex]{graphicx}
\else 
  \usepackage[dvips]{epsfig,graphicx}
  \usepackage[dvips,
    pdftitle={\papertitle},
    pdfauthor={\authorslist},
    pdfsubject={Proceedings of the 28th International Conference on Digital Audio Effects (DAFx25)},
    colorlinks=false, 
    bookmarksnumbered, 
    pdfstartview=XYZ 
  ]{hyperref}
\fi
\usepackage[hypcap=true]{caption}
\usepackage{cleveref}

\title{\papertitle}

\affiliation{
\paperauthorA \qquad \paperauthorB \,
\sthanks{This work was funded by UKRI and EPSRC as part of the ``UKRI CDT in Artificial Intelligence and Music'', under grant EP/S022694/1.}
}
{\href{https://dafx24.surrey.ac.uk}{Centre for Digital Music} \\ Queen Mary University of London, UK 
\\
{
\tt
\href{mailto:r.diazfernandez@qmul.ac.uk}{r.diazfernandez} \quad
\href{mailto:mark.sandler@qmul.ac.uk}{mark.sandler@qmul.ac.uk}}
}

\begin{document}
\ifpdf 
  \DeclareGraphicsExtensions{.png,.jpg,.pdf}
\else  
  \DeclareGraphicsExtensions{-eps-converted-to.pdf}
\fi

\makeatletter
\def\endthebibliography{%
  \def\@noitemerr{\@latex@warning{Empty `thebibliography' environment}}%
  \endlist
}
\makeatother

\maketitle

\begin{abstract}
\begin{sloppypar}
Modal methods for simulating vibrations of strings, membranes, and plates are widely used in acoustics and physically informed audio synthesis. However, traditional implementations, particularly for non-linear models like the von Kármán plate, are computationally demanding and lack differentiability, limiting inverse modelling and real-time applications. We introduce a fast, differentiable, GPU-accelerated modal framework built with the JAX library, providing efficient simulations and enabling gradient-based inverse modelling. Benchmarks show that our approach significantly outperforms CPU and GPU-based implementations, particularly for simulations with many modes. Inverse modelling experiments demonstrate that our approach can recover physical parameters, including tension, stiffness, and geometry, from both synthetic and experimental data. Although fitting physical parameters is more sensitive to initialisation compared to other methods, it provides greater interpretability and more compact parameterisation. The code is released as open source to support future research and applications in differentiable physical modelling and sound synthesis.
\end{sloppypar}
\end{abstract}

\section{Introduction}
\label{sec:intro}

The accurate modelling of vibrating structures, such as strings, membranes, and plates, is crucial to many areas, including acoustics, musical instrument design, and physically based audio synthesis. These systems are governed by partial differential equations (PDEs) derived from first principles. However, solving these PDEs analytically is possible only in a few specific cases. In practical settings, we therefore rely heavily on numerical approaches, each with distinct strengths and weaknesses.

Finite-difference time-domain (FDTD) methods are often used due to their explicit numerical formulation, discretising both time and space with finite differences. While intuitive and versatile, FDTD simulations often face significant limitations, such as numerical dispersion, restrictive stability conditions, and computational inefficiency, particularly for high-resolution applications.

Modal methods provide a different perspective by projecting the problem onto a carefully chosen set of orthogonal basis functions, transforming the PDE into a collection of simpler ordinary differential equations (ODEs). This approach offers advantages by naturally capturing the vibration modes of the system and reducing computational complexity. However, a significant challenge with modal methods is identifying appropriate basis functions, as these are dependent on the domain geometry and boundary conditions. Since analytical bases are available only for a limited number of simple cases, extending modal approaches to more complex scenarios remains difficult.  

Additionally, traditional numerical methods, including most modal approaches, are generally implemented without automatic differentiation, complicating the computation of gradients with respect to physical parameters. This limits their practical use for modern inverse modelling and gradient-based parameter estimation tasks, which are increasingly common in physically informed machine learning and audio synthesis. Furthermore, existing implementations typically run on CPUs, making them computationally inefficient for larger or non-linear problems, such as simulations involving the von Kármán plate model.

To address these challenges, this paper introduces a fast and differentiable GPU-accelerated framework for modal simulations of strings, membranes, and plates. Our implementation leverages the JAX library to enable efficient computation and automatic differentiation, making gradient-based optimisation directly accessible for parameter estimation, inverse modelling, and sound synthesis. 

\begin{itemize}
\item \textbf{Scalable computational efficiency:} Our GPU-based implementation significantly outperforms existing CPU and GPU implementations, especially for simulations involving many modes, making large-scale dataset generation and real-time applications feasible.

\item \textbf{Unified and accessible implementation of multiple models:} Our framework provides a single, user-friendly Python interface to simulate strings, membranes, and plates, including non-linear models like the von Kármán, Berger and Kirchoff-Carrier models.

\item \textbf{Physically interpretable inverse modelling:} We demonstrate inverse modelling by optimizing directly for physical parameters (e.g., tension, stiffness, geometry), rather than traditional spectral parameters (poles and zeros). Although this adds complexity due to increased constraints and sensitivity to initialization, it significantly improves interpretability and compactness of representation.
\end{itemize}
The open-source implementation provided with this paper connects classical physics-based modelling with modern automatic differentiation and GPU acceleration techniques. It offers audio researchers and practitioners an accessible, efficient, and interpretable tool for real-time synthesis, instrument design, and physically informed machine learning.

\section{Related Work}
\label{sec:related}

The simulation and modelling of strings, membranes, and plates has been extensively studied in acoustics and musical instrument research. Notably, the work of Bilbao~\cite{bilbao_numerical_2009} provides a comprehensive numerical treatment of such systems, primarily focusing on finite difference time-domain (FDTD) approaches. These methods discretise space and time, approximating derivatives using finite differences, and have been widely applied to the simulation of musical instruments and other acoustical systems. While FDTD methods are flexible and conceptually straightforward, they suffer from numerical dispersion and have restrictive stability conditions (e.g., CFL) that demand very small time steps. These constraints make high-frequency simulations or simulations involving large spatial domains computationally expensive. Recently, techniques like the Scalar Auxiliary Variable (SAV) method have also been introduced, primarily within FDTD frameworks, to improve the efficiency and stability of handling non-linear terms by avoiding implicit equations~\cite{bilbao_realtime_2023,wang_realtime_2023}.

Alternatively, modal approaches have been explored in different ways. Pseudospectral methods, which exploit Fourier decompositions, have been used for the simulation of the tension modulated string~\cite{bilbao_modaltype_2004}. Another significant modal approach is the Functional Transformation Method (FTM), introduced by Trautmann and Rabenstein~\cite{trautmann_functional_2003}. This technique relies on analytical spatial modes and the Laplace transform. By applying both transformations, the PDE becomes a set of transfer functions, which are then discretised and used in a numerical time integration scheme. However, analytical modal expansions exist only for specific geometries and boundary conditions. Subsequent works have focused on deriving analytical modes for more complex boundary conditions~\cite{schafer_continuous_2017}. In parallel, efficient tension-modulated modal models for strings and membranes have also been developed in the works of Avanzini and Marogna~\cite{avanzini_efficient_2012,avanzini_modular_2010,marogna_physicallybased_2009}.

In the specific context of plates, the non-linear von Kármán model has been extensively studied through modal formulations. The work of Duccheschi and Touzé~\cite{ducceschi_modal_2015,ducceschi_simulations_2015} addresses this non-linear model, particularly for the simply supported boundary condition. While the transverse (out-of-plane) vibration modes can be derived analytically for some cases, accurately capturing the non-linear coupling with in-plane modes typically requires a Galerkin type approach. Such methods select appropriate basis functions for projection onto a system of ODEs with coupling coefficients, enabling efficient numerical integration despite increased complexity due to non-linear mode coupling.

In recent years, differentiable and neural-based approaches have emerged, primarily targeting rigid-body simulations~\cite{clarke_diffimpact_2022,jin_diffsound_2024,diaz_rigidbody_2023}. Although most of these methods are primarily data-driven (rather than physics-based), they often leverage principles similar to classical modal expansions. In~\cite{diaz_efficient_2024}, non-linear modes are computed using gradient descent, and these modes are then evolved linearly, with a second neural network refining the dynamics to account for non-linear interactions. Similarly, Lee~\cite{lee_differentiable_2024} proposes using analytical (or alternatively learnt) modes as a basis, modulating them with neural networks to model pitch glide and non-linear coupling effects. Such hybrid approaches highlight the potential benefits of integrating physics-based modal techniques with neural methods.

Our framework remains entirely physics-based and does not depend on neural networks for modelling. Instead, we implement traditional modal formulations in a unified, differentiable manner, leveraging automatic differentiation and GPU acceleration to improve computational efficiency. Due to its differentiability and modular design, the framework can readily be combined with neural networks if desired, enabling hybrid physics-based and data-driven approaches. By offering accessible implementations of tension-modulated string, membrane, and von Kármán plate models, our method simplifies gradient-based inverse modelling and efficient differentiable simulations, effectively bridging classical modal techniques with modern automatic differentiation methods.

\section{Background}
\label{sec:background}

Assuming uniform material properties, the governing equations for the tension-modulated string, membrane, and von Kármán plate with tension can be written as:
\begin{equation}
\label{eq:general}
\rho \ddot{w} + \left(d_1 + d_3 \Delta\right)\dot{w} + (D \Delta \Delta - T_0 \Delta) w = f_{\text{ext}} - f_{\text{nl}},
\end{equation}
where $w$ is the transverse displacement, $\rho$ is the material density, $d_1$ and $d_3$ are the damping coefficients, $D$ is the bending stiffness, $T_0$ is the initial tension, $f_{\text{ext}}$ is the external force and $f_{\text{nl}}$ is the non-linear term. $\Delta$ and $\Delta \Delta$ stand for the Laplacian and biharmonic operators, respectively.

For the modal approach, a suitable set of eigenfunctions must be found by solving the spatial eigenvalue problem:
\begin{equation}
\label{eq:eigenvalue}
\Delta \Phi_{\mu} = -\lambda_{\mu} \Phi_{\mu},
\end{equation}
where $\Phi_{\mu}$ is the $\mu$-th eigenfunction and $\lambda_{\mu}$ is the corresponding eigenvalue. 

In general, analytical solutions for these eigenpairs are known only for simple geometries and boundary conditions. For more complex scenarios, numerical methods such as the Finite Element Method (FEM) or other Galerkin-type approaches are typically employed.

In the linear case, the non-linear term $f_{\text{nl}}$ in ~\cref{eq:general} vanishes and the equation can be solved as a system of uncoupled ODEs. In the tension-modulated case, that is the Kirchoff-Carrier model in the case of a string and Berger model in the case of a membrane, the non-linear term arises due to the tension caused by the deformation of the string or membrane~\cite{avanzini_efficient_2012,trautmann_sound_2000,bilbao_numerical_2009}:
\begin{equation}
\label{eq:tension}
f_{\text{nl}} = \tau \int_\mathcal{D} ||\nabla w||^2 d\mathbf{x},
\end{equation}
where $\mathcal{D}$ is domain length or area and $\tau$ is a scalar dependent on the physical parameters of the string or membrane: 
\begin{equation}
    \tau_{\text{string}} = \frac{E A}{2 L}, \quad \tau_{\text{membrane}} = \frac{E h}{2 L_x L_y (1 - \nu^2)}
\end{equation}
where $E$ is the Young's modulus, $A$ is the cross-sectional area, $L$ is the length of the string, $h$ is the thickness, $L_x$ and $L_y$ are the length and width, and $\nu$ is the Poisson's ratio.

For the von Kármán plate, nonlinearities emerge from the coupling between transverse and in-plane modes~\cite{ducceschi_modal_2015,ducceschi_simulations_2015}:
\begin{align}
\label{eq:plate}
f_{nl} &= -\mathcal{L}(w, \varphi), \\
\Delta \Delta \varphi &= -\frac{Eh}{2} \mathcal{L}(w, w),
\end{align}
where $\mathcal{L}$ is the von Kármán operator defined as:
\begin{equation}
\label{eq:von_karmen}
\mathcal{L}(f, g)=\Delta f \Delta g-\nabla \nabla f: \nabla \nabla g,
\end{equation}
and $\varphi$ is the Airy stress function.

\subsection{Modal expansion of the non-linear term}
\label{subsec:modal_expansion}

For tension-modulated strings and membranes, the non-linear term expanded in modal coordinates is given by:
\begin{equation}
\label{eq:modal_string_membrane}
\bar{f}_{nl} = \lambda_{\mu} \tau \sum_{\mu} \frac{\lambda_{\mu} q_{\mu}^2}{||\Phi_{\mu}||^2} q_{\mu},
\end{equation}
where $q_{\mu}$ is the amplitude of the $\mu$-th modal coordinate.

For the von Kármán plate, both the transverse modes $\Phi_{\mu}$ and in-plane modes $\Psi_{\mu}$ must be considered. The non-linear modal expansion then becomes:
\begin{equation}
\label{eq:nl_plate}
\bar{f}_{nl} = \frac{E}{2 \rho} \sum_{p, q, r}^n \frac{H_{q, r}^n C_{p, n}^s}{\zeta_n^4} q_p q_q q_r,
\end{equation}
where $H_{q, r}^n$ and $C_{p, n}^s$ are third-order tensors with the coupling coefficients obtained from the projections between the modes:
\begin{align}
\label{eq:coupling_coefficients}
H_{i, j}^k & =\frac{\int_S \Psi_k \mathcal{L}\left(\Phi_i, \Phi_j\right) \mathrm{d} S}{\left\|\Psi_k\right\|\left\|\Phi_i\right\|\left\|\Phi_j\right\|}, \\
C_{i, j}^s & =\frac{\int_S \Phi_s \mathcal{L}\left(\Phi_i, \Psi_j\right) \mathrm{d} S}{\left\|\Phi_s\right\|\left\|\Phi_i\right\|\left\|\Psi_j\right\|},
\end{align}
where $\Psi$ and $\zeta^4$ are the eigenfunctions and eigenvalues obtained from the separate eigendecomposition of the Airy stress function $\varphi$. For a more detailed derivation of the modal expansion of the non-linear term, we refer the reader to~\cite{ducceschi_modal_2015,ducceschi_simulations_2015}.

\subsection{Integration in time}

Expressing Equation~\eqref{eq:general} in modal coordinates yields a system of ODEs describing individual modes as damped harmonic oscillators:
\begin{equation}
    \label{eq:modal_oscillator}
    \ddot{q}_{\mu} + 2\gamma_{\mu}\dot{q}_{\mu} + \omega_{\mu}^2 q_{\mu} = \bar{f}_{\text{ext},\mu} - \bar{f}_{\text{nl},\mu},
\end{equation}
where the coefficients are given by:
\begin{align}
\omega_{\mu}^2 &= \frac{D\lambda_{\mu}^2 + T_0\lambda_{\mu}}{\rho},\\[5pt]
\gamma_{\mu} &= \frac{d_1 + d_3\lambda_{\mu}}{2\rho}.
\end{align}

\begin{figure}[h]
\centering
\includegraphics[width=\columnwidth, trim=5 5 5 5, clip]{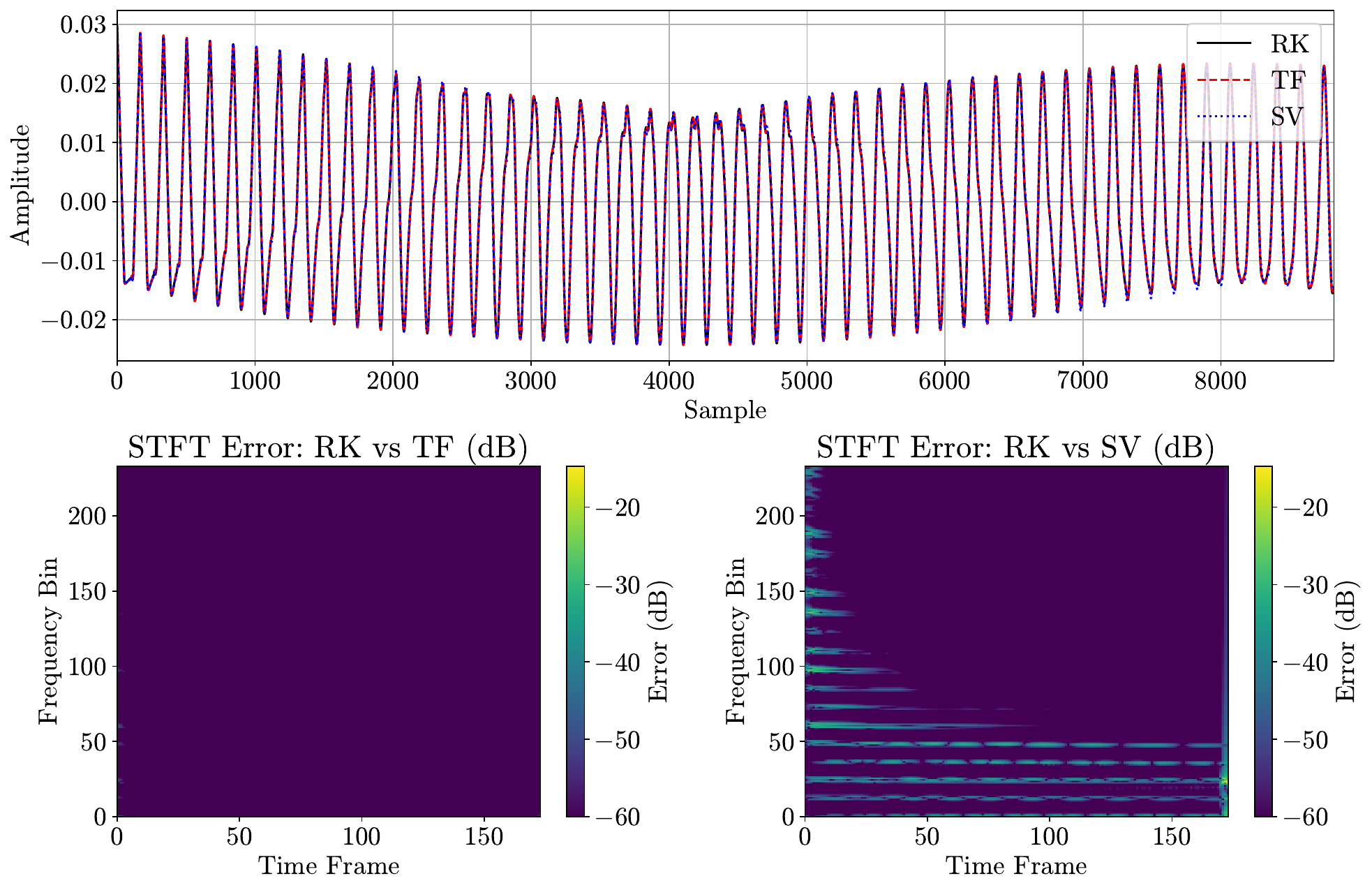}
\caption{Error comparison between numerical integration methods. \textbf{Top}: Simulation of the tension-modulated string from initial conditions using DOP853 (RK), the FTM-based method (TF), and the Störmer-Verlet method (SV). \textbf{Bottom}: Error computed as the magnitude STFT difference between each method and the reference DOP853 integrator. The simulation was performed for a tension-modulated string with 40 modes.}
\label{fig:stft_comparison}
\end{figure}

One possible solution to the ODEs in~\Cref{eq:modal_oscillator}, following the FTM approach, is to use the Laplace transform and the impulse invariance discretisation to produce discrete transfer functions:
\begin{equation}
\label{eq:transfer_function}
H_{\mu}(z) = 
\frac{
b_{\mu, 1} z
}
{z^2 + a_{\mu, 1}z + a_{\mu,2}} + nl_{\mu}(z),
\end{equation}
%
\begin{align}
a_{\mu, 1} &= -2 e^{-\gamma_{\mu}T}\cos(\tilde{\omega}_{\mu}T), \\
a_{\mu, 2} &= e^{-2\gamma_{\mu}T},\\
b_{\mu, 1} &= \frac{\rho^{-1}\sin(\tilde{\omega}_{\mu}T)}{\tilde{\omega}_{\mu}} e^{-\gamma_{\mu}T},   
\end{align}
with \(\tilde{\omega}_{\mu} = \sqrt{\omega_{\mu}^2 - \gamma_{\mu}^2}\), which yields the following update scheme for all the modes:
\begin{equation}
\label{eq:df_update}
\mathbf{q}^{n+1} = \mathbf{a_1} \odot \mathbf{q}^n + \mathbf{a_2} \odot \mathbf{q}^{n-1} + \mathbf{b_1} (\mathbf{\bar{f}}_{ext}^n - \mathbf{\bar{f}}_{nl}^n),
\end{equation}
A similar second-order approach is to use St\"ormer-Verlet~\cite{hairer_geometric_2003} directly on ~\cref{eq:modal_oscillator}. Discretising the ODE in time using centered finite differences we obtain the following update rule:
\begin{equation}
\label{eq:modal_oscillator_update}
\mathbf{q}^{n+1} = \mathbf{g} \odot \mathbf{q}^n + \mathbf{p} \odot \mathbf{q}^{n-1} + \mathbf{r} \odot (\mathbf{\bar{f}}_{ext}^n - \mathbf{\bar{f}}_{nl}^n),
\end{equation}
where $\mathbf{g},\mathbf{p}, \mathbf{r}$ are vectors of coefficients obtained from the time discretisation for each $\mu$-th mode:
\begin{align}
    r_\mu &= \frac{2T^2}{2 + 2 \gamma_\mu T} \\
    g_\mu &= r_\mu \left(\frac{2}{T^2} - \omega_\mu^2\right) \\
    p_\mu &= r_\mu \left(-\frac{1}{T^2} + \frac{2\gamma_\mu}{2T}\right),
\end{align}
In both cases, $\mathbf{\bar{f}}_{\text{ext}}^n$ and $\mathbf{\bar{f}}_{\text{nl}}^n$ denote the modal coordinate vectors of external forces and non-linear terms, respectively, and $T$ denotes the sampling period. While higher-order integration schemes could be employed for improved accuracy, we choose second-order methods due to their computational efficiency. Figure~\ref{fig:stft_comparison} compares the numerical errors of these second-order schemes against a reference solution computed with an eighth-order Runge-Kutta integrator (DOP853).

\section{Method}
\label{sec:method}

\subsection{Implementation}
\label{subsec:implementation}

We implemented the framework in Python using JAX~\cite{bradbury_jax_2018} which enables efficient, GPU-accelerated simulations with automatic differentiation. The full source code for all experiments, benchmarks, and figures presented in this paper is available at: \url{https://github.com/rodrigodzf/jaxdiffmodal} as well as an accompanying website for documentation and usage of the framework.

The open source repository includes utilities for computing analytical modes and eigenvalues of fixed strings, membranes and plates. For plates, we provide a Python implementation of the approach described in~\cite{ducceschi_simulations_2015} to compute non-linear coupling coefficients\footnote{\label{vkgong}\url{https://vkgong.ensta-paris.fr}}. To support arbitrary boundary conditions, we use the \texttt{magpie-python} library\footnote{\url{https://github.com/Nemus-Project/magpie-python}} together with a reimplementation for processing numerical modes based on\footnote{\label{vkplate_repo}\url{https://github.com/Nemus-Project/VKPlate}}.

The framework includes differentiable implementations of the time integration schemes discussed in Section~\ref{sec:background}, along with tools for modal to physical domain transformations, transfer function computation, and visualisation. All components are modular and designed to support easy experimentation and extension.

\subsection{Benchmarks}
\label{sec:benchmarks}

\begin{figure}[h]
\centering
\includegraphics[width=\columnwidth, trim=5 0 0 10pt, clip]{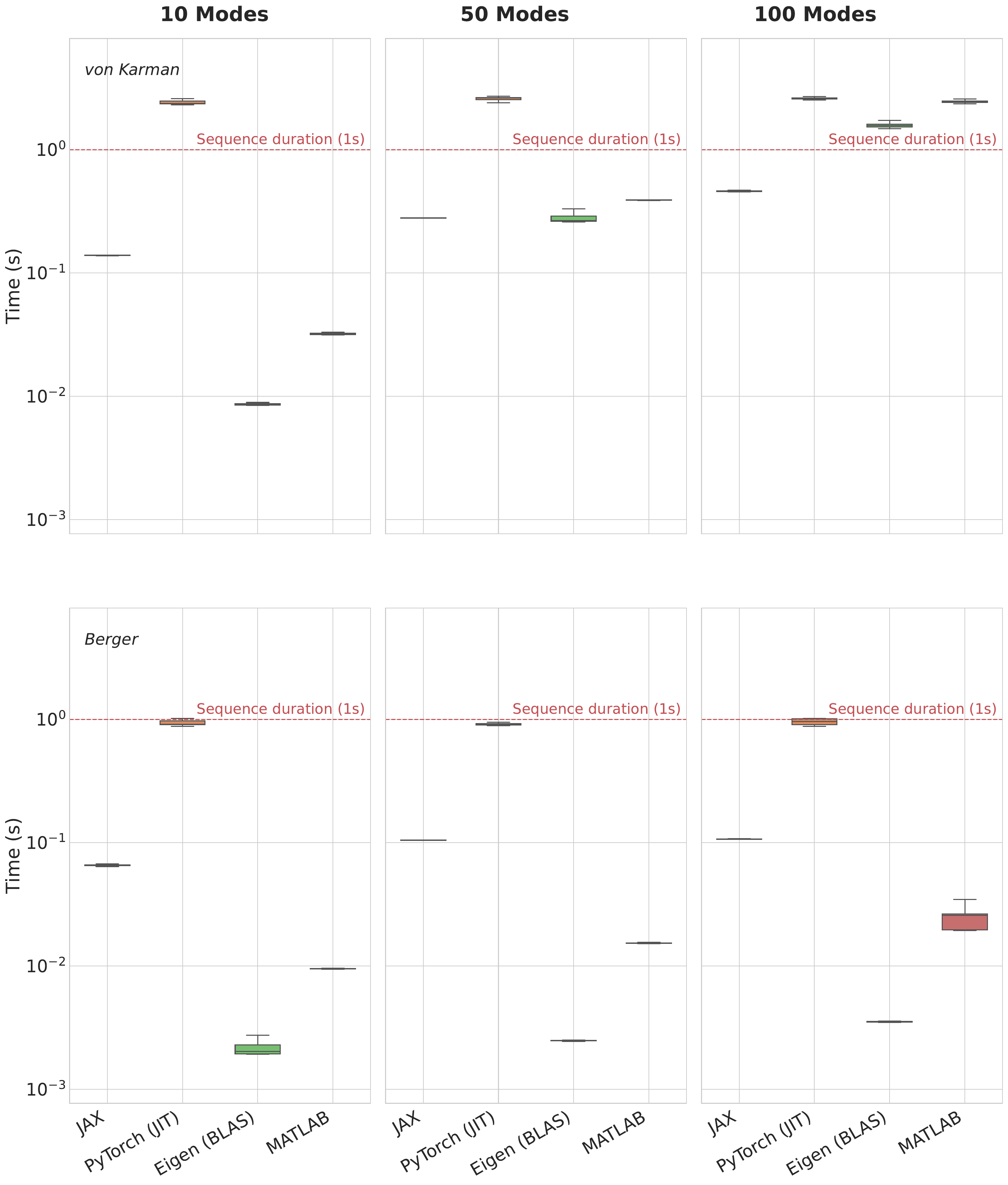}
\caption{Benchmark comparison of different implementations. \textbf{Top:} von Kármán plate model, \textbf{Bottom:} tension modulated plate model (Berger). Performance is measured for simulations of 10, 50, and 100 modes, each running for 1 second at 44,100 Hz. A dotted line is shown to indicate real-time performance, above the line is slower than real-time. Results are displayed as box plots obtained from 50 repeated runs per configuration.}
\label{fig:benchmarks}
\end{figure}

A key motivation for this work is to develop a high-performance simulation framework for non-linear models, particularly aimed at efficient large-scale dataset generation. To evaluate computational performance, we benchmark our St\"ormer-Verlet time integration approach (computationally equivalent to the discretised FTM method) for the von Kármán plate model implemented in JAX against three alternative implementations: a MATLAB version, an optimised C++ implementation using the Eigen library with BLAS support, and a JIT-compiled PyTorch implementation. This comparison is particularly challenging due to the tensor contraction (in \Cref{eq:nl_plate}) required at each time step. As shown in Figure~\ref{fig:benchmarks}, while our JAX-based approach is moderately slower than the MATLAB and C++ implementations for smaller mode counts ($\lessapprox 50$), it demonstrates better efficiency as the number of modes increases. It also substantially outperforms the optimised PyTorch implementation running on the GPU. Notably, for the von Kármán plate model with approximately 100 modes, our implementation is the only one capable of achieving roughly twice real-time performance for a one-second simulation at a 44,100 Hz sampling rate.

All benchmarks were performed on a system with an AMD Ryzen 9 5900X CPU, an NVIDIA GeForce RTX 3090 GPU, and 64GB of RAM, using single-precision floating point.

We also developed a custom CUDA kernel implementation, but it is not included in the benchmark results as it proved unexpectedly slower than the other approaches.

\subsection{Experiments}
\label{subsec:experiments}

We present three experiment groups designed to evaluate the differentiability and effectiveness of our framework for inverse modelling. These experiments assess the ability of our method to identify and optimise physical parameters for strings and plates using both synthetic simulations and real-world measurements (mainly for linear cases).

\subsubsection{Loss functions and setup}
\label{subsubsec:loss_functions_and_setup}

\begin{figure}[h]
\centering
\includegraphics[width=\columnwidth]{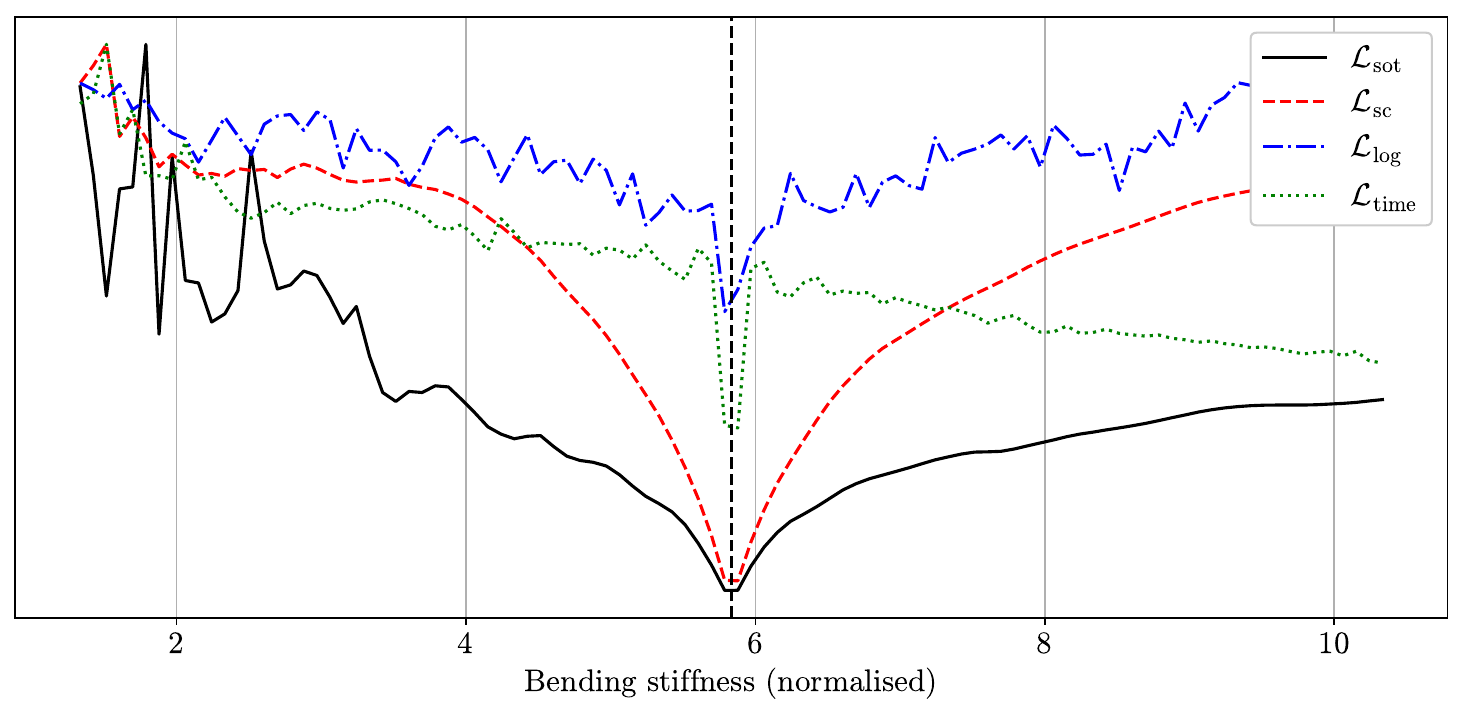}
\caption{Loss landscapes with respect to the normalized bending stiffness $\hat{D}$. Three losses are considered: log-magnitude ($\mathcal{L}_{\text{log}}$), spectral convergence ($\mathcal{L}_{\text{sc}}$), and spectral optimal transport ($\mathcal{L}_{\text{sot}}$). The time-domain MSE ($\mathcal{L}_{\text{time}}$) is included for comparison. The vertical dotted line indicates the optimal value of $\hat{D}$ for the target response. The losses are scaled for better visualization.}
\label{fig:loss_landscape_time}
\end{figure}

For inverse modelling of physical parameters, we employ a composite loss function based on spectral differences between predicted and target spectral magnitudes. This loss function, common across all experiments, combines three complementary spectral metrics:
\begin{align}
\label{eq:loss_functions}
\mathcal{L}_{\text{log}} &= ||\log (Y + \epsilon) - \log (\hat{Y} + \epsilon)||_1, \\
\mathcal{L}_{\text{sc}} &= \frac{||Y - \hat{Y}||_F}{||Y||_F}, \\
\mathcal{L}_{\text{sot}} &= \frac{1}{N} \sum_{i=0}^{N} \mathcal{W}_1\left(Y_i, \hat{Y}_i\right), \\
\mathcal{L}_{\text{total}} &= \alpha \mathcal{L}_{\text{log}} + \beta \mathcal{L}_{\text{sc}} + \eta \mathcal{L}_{\text{sot}},
\end{align}
where $x$ and $\hat{x}$ represent the time-domain target and predicted signals, respectively, with $Y = |\text{STFT}(x)|$ and $\hat{Y} = |\text{STFT}(\hat{x})|$ denoting their corresponding STFT magnitude spectrograms. The terms $Y_i$ and $\hat{Y}_i$ refer to individual (normalised) time frames of the spectrograms used in the spectral optimal transport loss $\mathcal{L}_{\text{sot}}$~\cite{torres_unsupervised_2024} via the Wasserstein-1 distance $\mathcal{W}_1$. The constant $\epsilon$ ensures numerical stability in the logarithmic term. The weights $\alpha$, $\beta$, and $\eta$ are calibrated for each experiment to balance the contribution of each loss component. For linear cases, we also compute a slight variation of these losses where the STFT is replaced by the sampled magnitude frequency response, which provides a more direct measure of the systems transfer function (available in this specific case). Although $\mathcal{L}_{\text{sot}}$ helps avoid poor local minima, the overall spectral loss landscape remains non-convex, as illustrated in Figure~\ref{fig:loss_landscape_time}, even for a single parameter.

We perform optimisation using the Adam optimiser with a one-cycle cosine learning rate schedule. For linear cases, we run 15,000 optimisation steps, while for non-linear cases, we conduct 1,000 steps with 100 parallel random initializations to improve convergence.

Some of the parameters are optimised in a \textit{density-normalised} form: $\hat{D} = D / \rho$ and $\hat{T}_0 = T_0 / \rho$. For damping parameters, we also directly optimise the $\gamma_{\mu}$ coefficients in some of the experiments to enhance model flexibility when fitting to real-world data.

By optimising the parameters $\hat{D}$ and $\hat{T}_0$, we implicitly optimise the modal frequencies while preserving the dispersion relation and eigenvalues determined by the geometry and boundary conditions (solutions to ~\cref{eq:eigenvalue}). This approach differs fundamentally from methods that optimise modes independently of physical constraints~\cite{diaz_rigidbody_2023}.

\subsubsection{String}
\label{subsubsec:string}

\begin{figure}[ht!]
\captionsetup[subfloat]{}

\subfloat[]{%
    \includegraphics[width=\columnwidth]{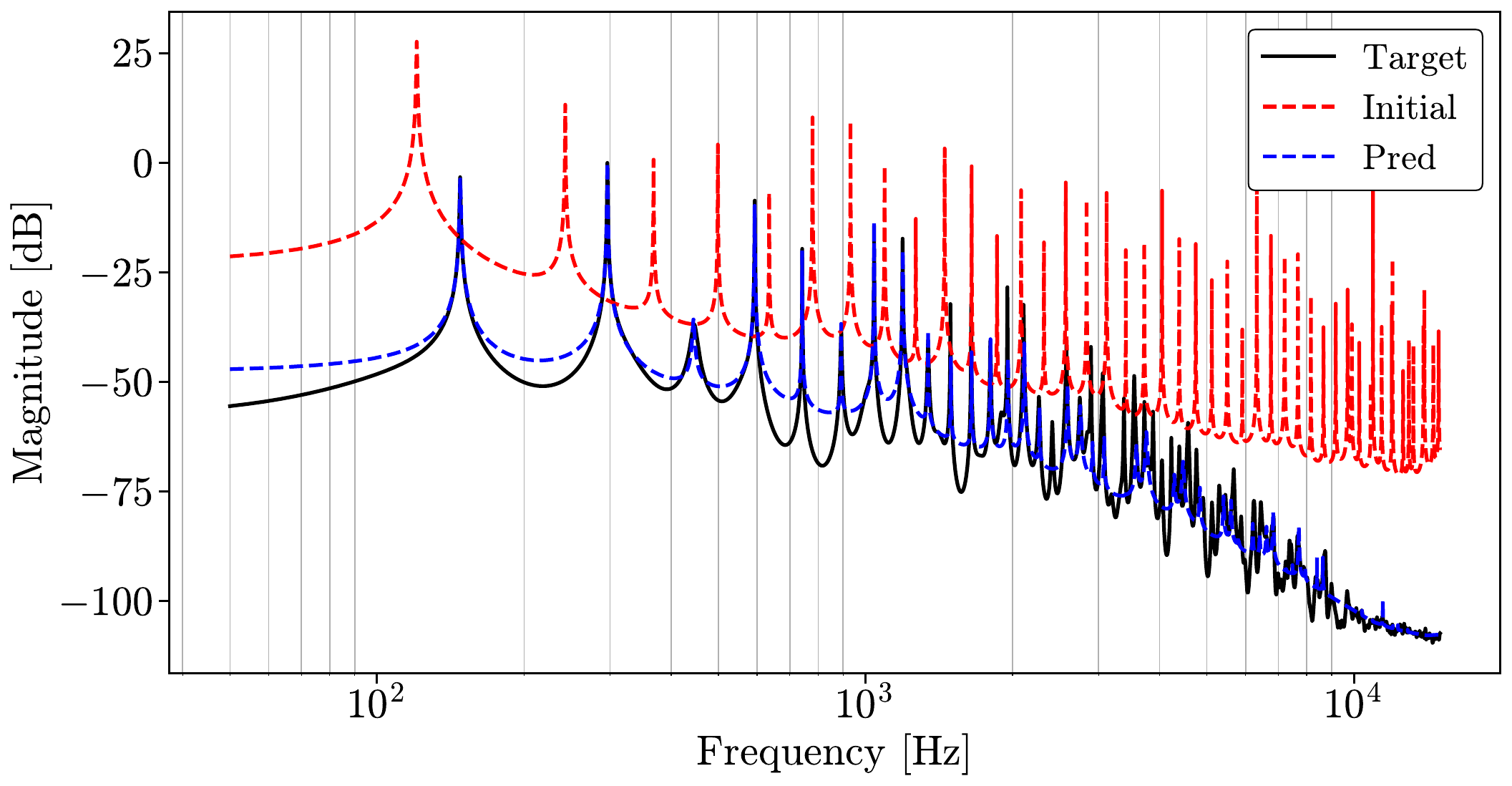}%
    \label{fig:string_fit}%
}\vspace{-10pt}

\subfloat[]{%
    \includegraphics[width=\columnwidth]{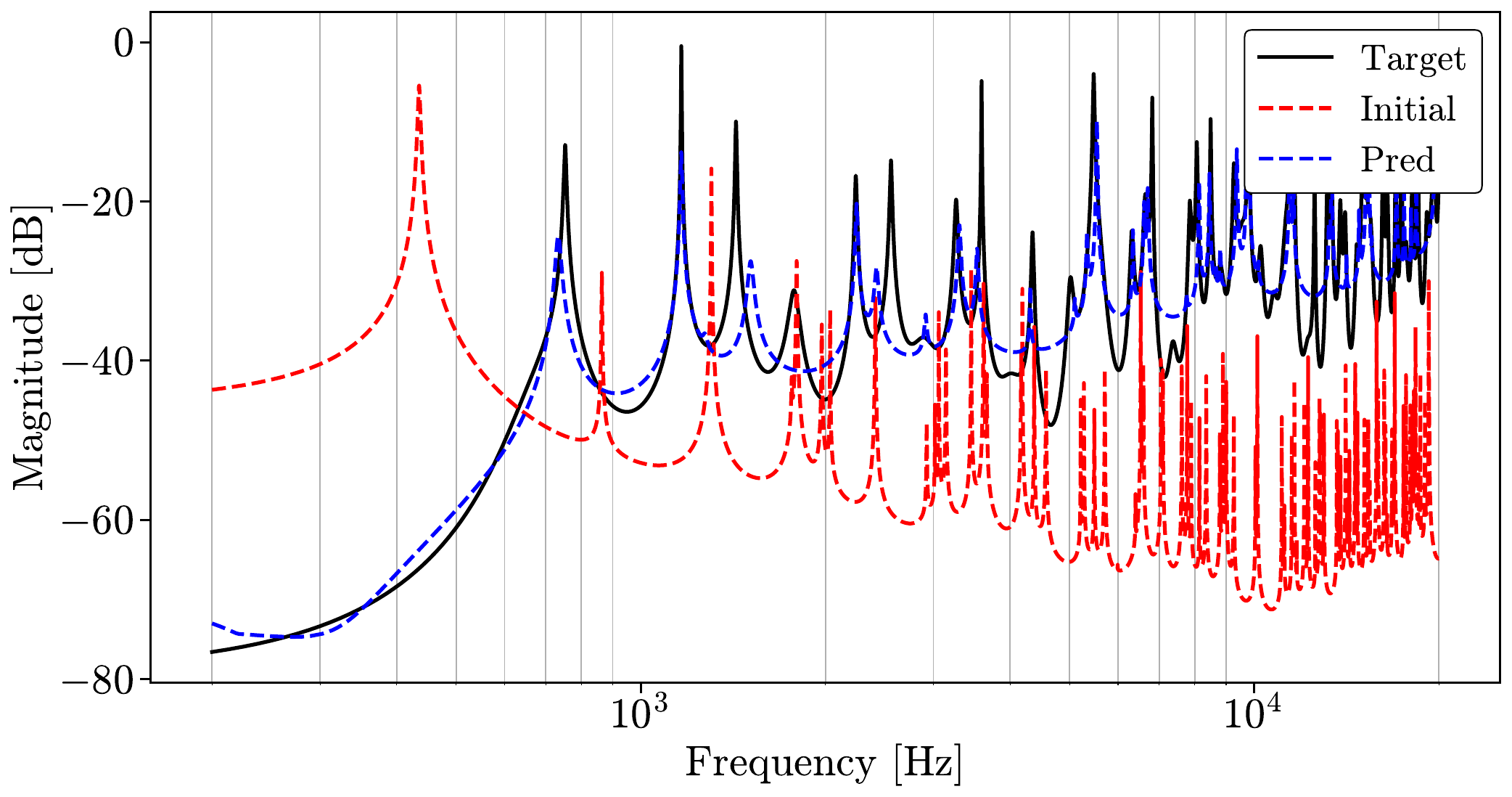}%
    \label{fig:plate_fit}%
}
\captionsetup{skip=5pt}

\caption{Inverse modelling experiments. (a) Semi-log plot of the fit to a real plucked string. (b) Semi-log plot of the fit to a real struck thick plate. (c) Spectrogram of the target and optimised responses of a synthetic simulation of the von Kármán plate model.}
\label{fig:inverse_modelling}
\end{figure}
    
We first consider the linear string model, focusing on parameter fitting for both synthetic and real string responses. Two optimisation approaches are used for this task.

In the first approach, and in particular for the real-world scenarios, we extract the spectral envelope from the target response using high-order Linear Predictive Coding (LPC), though alternative methods could be employed. We then sample the LPC coefficients in frequency using the Bark scale. On the model side, we compute the transfer function of the linear system using~\cref{eq:transfer_function}, based on a set of parameters to be optimised. The predicted response is sampled in the same way as the LPC envelope. We then compute the loss using the functions defined in~\cref{eq:loss_functions}, replacing the STFT magnitude with the Bark-sampled magnitude response. This provides a simple and efficient route for inverse modelling in the linear case, avoiding time-domain simulation.~\Cref{fig:string_fit} shows the semi-log plot of the fit to a real plucked string.

In the second approach, we simulate the string response in time by solving the linear ODE directly. This is done through recursive multiplication of the discrete complex eigenvalues derived from the dispersion relation $s_{\mu\pm} = \gamma_\mu \pm i \tilde{\omega}_\mu$. This can be efficiently implemented using the prefix-sum algorithm~\cite{blelloch_prefix_1990} allowing fast evaluation across modes. It is also possible to use the solving schemes (\cref{eq:modal_oscillator_update,eq:df_update}), however for the linear case this is not necessary.

We fit the parameters of a single recording of a real plucked string from the IDMT-SMT-Guitar dataset~\cite{kehling_idmtsmtguitar_2023}. For this experiment, we assume the string is fixed at both ends and that nonlinear effects (e.g. pitch glide) are minimal.

While optimisation converges more quickly in the synthetic case, the real-world scenario presents greater challenges. This increased complexity arises from the need to account for not only the PDE parameters but also the external force excitation, the acoustic transfer function, measurement noise, and other effects.

We extend the model by optimising an additional parameter $b_{\mu,2}$ (in~\Cref{eq:transfer_function}) for each mode, to address the unknown initial conditions. Furthermore, instead of projecting the solution back to the physical domain using analytical mode shapes, we simply optimise a set of abstract weights for each mode. These weights account for effects such as the acoustic transfer and microphone placement, even if they are not physically interpretable.

\subsubsection{Plate}
\label{subsubsec:plate}

\begin{figure*}[h]
\centering
\includegraphics[width=\textwidth]{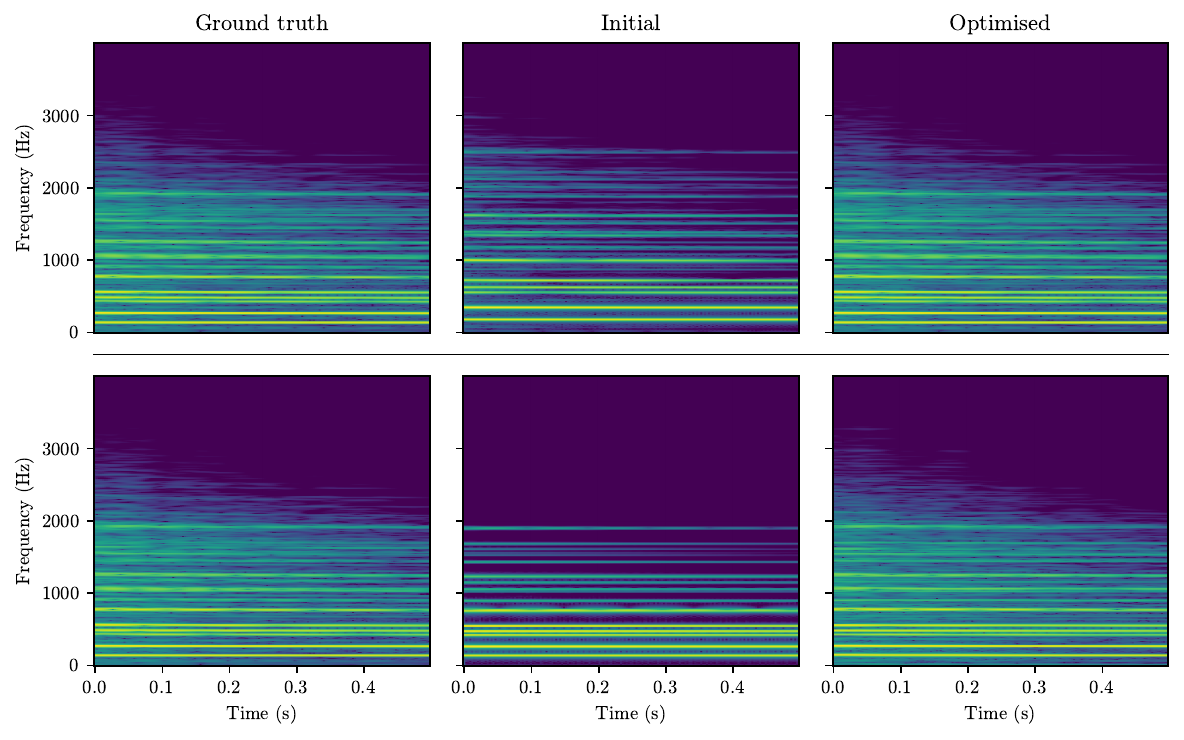}
\caption{Spectrograms of the target and optimised forced responses from simulations of the von Kármán plate model. \textbf{Top}: Matching the response at a single point on the plate, by optimising only the normalised bending stiffness ($\hat{D}_{\text{target}} = 5.8328$, $\hat{D}_{\text{initial}} = 10$, $\hat{D}_{\text{optimised}} = 5.8329$). \textbf{Bottom:} Matching the response by optimising only the coupling coefficients $H$, initialised with random values from a normal distribution.}
\label{fig:spectrograms}
\end{figure*}

We apply a similar procedure to the linear plate model simulations, examining both synthetic and real scenarios. The real-world case uses a single hammer strike recording from a thick rigid plate taken from the RealImpact dataset~\cite{clarke_realimpact_2023}.

In the synthetic non-linear case, we simulate the displacement of the plate at a fixed point in the domain. We then optimise the parameters by propagating gradients through the time integrator to match this response (i.e., using BPTT). This type of optimisation is more sensitive to parameter initialisation of parameters and is prone to getting trapped in local minima. Even when optimising a single parameter, such as $\hat{D}$, the loss landscape exhibits many local minima (\Cref{fig:loss_landscape_time}). This behaviour is also consistent with previously observed challenges in the optimisation of frequencies~\cite{turian_im_2020a}. In addition, care must be taken to ensure that all parameters remain within numerically stable ranges. In particular, the damping coefficients $\gamma_{\mu}$ must be constrained to ensure all poles lie in the left half of the Laplace domain to avoid numerical instability. To address the challenge of optimising the bending stiffness and avoiding poor local minima, we adjust the weighting of $\mathcal{L}_{\text{total}}$ to emphasise the spectral optimal transport and spectral convergence losses. Since the optimisation using automatic differentiation runs is fast (for approximately 0.1 seconds at 44,100 Hz), we can efficiently explore multiple random initialisations in parallel~\cite{dellasanta_automatic_2024}.

We also experiment with the optimisation of the coupling coefficients $C$ and $H$ (in~\Cref{eq:coupling_coefficients}) of the von Kármán plate model, which are otherwise difficult to compute analytically or numerically. In the simply supported case, only the $H$ tensor needs to be optimised, since it satisfies the identity $H_{i, j}^k = C_{i, k}^j$~\cite{ducceschi_nonlinear_2014}. We initialise the tensor with random values drawn from a normal distribution and optimise it to match the simulated plate displacement at a fixed point in space. This optimisation is carried out over a sequence of 17,640 time steps (0.4 seconds at 44,100 Hz).

\Cref{fig:spectrograms} shows the spectrograms of the target, initial, and optimised forced responses from a synthetic von Kármán plate simulation. The results show that the optimisation successfully converges to the correct behaviour. Although the fit can be performed on a short segment (0.4 seconds), the optimised parameters remain valid and consistent for longer simulations.

\section{Conclusion}
\label{sec:conclusion}

In this paper, we have presented a fast, differentiable modal simulation framework for strings, membranes, and plates. By leveraging GPU acceleration through the JAX library, our method significantly outperforms traditional CPU-based implementations, enabling efficient simulation, dataset generation, and inverse modelling, even for non-linear models such as the von Kármán plate.

The benchmarks demonstrate that our implementation scales efficiently with an increasing number of modes, providing substantial performance gains compared to existing MATLAB, optimised C++, and GPU-based PyTorch (JIT) implementations. Furthermore, it supports effective gradient-based optimisation of physical parameters from both synthetic simulations and real-world measurements. We successfully recovered parameters such as tension, bending stiffness, damping, and geometry, including the coupling tensors of the non-linear von Kármán plate model.

However, directly optimising physical parameters rather than fitting more abstract representations such as damped exponentials or poles introduces significant challenges. The physical parameterisation adds complexity and non-linear constraints, making optimisation sensitive to initialisation and prone to convergence to poor local minima. Despite these challenges, the physical approach offers considerable benefits. First, the results are directly interpretable since the parameters correspond to physically meaningful quantities. Second, the representation is highly compact, capturing behaviour that would otherwise require many free parameters with just a few well-defined physical ones.

Despite the advantages, several limitations remain. The ill-posed nature of inverse modelling in this case needs careful initialisation and robust optimisation strategies are essential. While using multiple random initialisations helps mitigate these issues, a more structured approach to parameter initialisation and optimisation scheduling is needed to improve robustness and convergence. For example, a staged optimisation approach similar to that of~\cite{jin_diffsound_2024}, in which damping and frequencies are optimised at different stages, may improve convergence stability.

Looking ahead, we plan to explore several improvements and extensions. First, we aim to generalise the solver to support three-dimensional structures using a differentiable finite element method (FEM) solver. Second, we will reimplement a modal SAV method and the energy-conserving integration scheme in Python, both of which have been previously implemented in MATLAB. Third, the current coupling tensor computations could benefit from sparsity-aware techniques. Exploiting the inherent structure of these tensors through sparse matrix representations could lead to significant speedups for high mode count simulations.

In summary, our work bridges classical modal techniques with modern automatic differentiation methods, offering a highly efficient, differentiable, and physically interpretable framework. It enables new applications in audio synthesis, inverse modelling, and physically informed machine learning, while opening avenues for future research into differentiable physics-based modelling.

\section{Acknowledgments}

We would like to thank Alexis Mousseau and Matthew Hamilton for their helpful comments and support with the \texttt{VKPlate} and \texttt{magpie-python} projects. We also thank Carlos de La Vega for his insights during discussions on the non-linear models.

\bibliographystyle{IEEEbib}
\bibliography{references} 

\begin{thebibliography}{10}

\bibitem{bilbao_numerical_2009}
Stefan Bilbao,
\newblock {\em Numerical {{Sound Synthesis}}: {{Finite Difference Schemes}} and {{Simulation}} in {{Musical Acoustics}}},
\newblock Wiley, 1 edition, Oct. 2009.

\bibitem{bilbao_realtime_2023}
Stefan Bilbao, Craig Webb, Zehao Wang, and Michele Ducceschi,
\newblock ``Real-time gong synthesis,''
\newblock in {\em Proceedings of the 26th International Conference on Digital Audio Effects}, Sept. 2023, Proceedings of the International Conference on Digital Audio Effects, pp. 1--8.

\bibitem{wang_realtime_2023}
Zehao Wang, Stefan Bilbao, Tom Erbe, and Miller Puckette,
\newblock ``Real-time implementation of the {{Kirchhoff}} plate equation using finite-difference time-domain methods on {{CPU}},''
\newblock in {\em Proceedings of the 20th {{Sound}} and {{Music Computing Conference}}}. June 2023, pp. 354--361, {Sound and Music Computing Network}.

\bibitem{bilbao_modaltype_2004}
Stefan Bilbao,
\newblock ``Modal-{{Type Synthesis Techniques}} for {{Nonlinear Strings}} with an {{Energy Conservation Property}},''
\newblock in {\em Proceedings of the 7th {{International Conference}} on {{Digital Audio Effects}} ({{DAFx04}})}, 2004.

\bibitem{trautmann_functional_2003}
Lutz Trautmann and Rudolf Rabenstein,
\newblock ``Functional {{Transformation Method}},''
\newblock in {\em Digital {{Sound Synthesis}} by {{Physical Modeling Using}} the {{Functional Transformation Method}}}, Lutz Trautmann and Rudolf Rabenstein, Eds., pp. 95--187. Springer US, Boston, MA, 2003.

\bibitem{schafer_continuous_2017}
Maximilian Sch{\"a}fer and Rudolf Rabenstein,
\newblock ``A {{Continuous Frequency Domain Description}} of {{Adjustable Boundary Conditions}} for {{Multidimensional Transfer Function Models}},''
\newblock 2017.

\bibitem{avanzini_efficient_2012}
Federico Avanzini, Riccardo Marogna, and Bal{\'a}zs Bank,
\newblock ``Efficient synthesis of tension modulation in strings and membranes based on energy estimation,''
\newblock {\em The Journal of the Acoustical Society of America}, vol. 131, no. 1, pp. 897--906, Jan. 2012.

\bibitem{avanzini_modular_2010}
Federico Avanzini and Riccardo Marogna,
\newblock ``A {{Modular Physically Based Approach}} to the {{Sound Synthesis}} of {{Membrane Percussion Instruments}},''
\newblock {\em IEEE Transactions on Audio, Speech, and Language Processing}, vol. 18, no. 4, pp. 891--902, May 2010.

\bibitem{marogna_physicallybased_2009}
Riccardo Marogna and Federico Avanzini,
\newblock ``Physically-based synthesis of nonlinear circular membranes,''
\newblock in {\em Proc. {{Int}}. {{Conf}}. {{Digital Audio Effects}} ({{DAFx-09}})}, 2009.

\bibitem{ducceschi_modal_2015}
Michele Ducceschi and Cyril Touz{\'e},
\newblock ``Modal approach for nonlinear vibrations of damped impacted plates: {{Application}} to sound synthesis of gongs and cymbals,''
\newblock {\em Journal of Sound and Vibration}, vol. 344, pp. 313--331, May 2015.

\bibitem{ducceschi_simulations_2015}
Michele Ducceschi and Cyril Touz{\'e},
\newblock ``Simulations of nonlinear plate dynamics: {{An}} accurate and efficient modal algorithm,''
\newblock in {\em Proceedings of the 18th {{International Conference}} on {{Digital Audio Effects}}}. 2015, {Norwegian University of Science and Technology}.

\bibitem{clarke_diffimpact_2022}
Samuel Clarke, Negin Heravi, Mark Rau, Ruohan Gao, Jiajun Wu, Doug James, and Jeannette Bohg,
\newblock ``Diffimpact: {{Differentiable}} rendering and identification of impact sounds,''
\newblock in {\em Conference on {{Robot Learning}}}. 2022, pp. 662--673, PMLR.

\bibitem{jin_diffsound_2024}
Xutong Jin, Chenxi Xu, Ruohan Gao, Jiajun Wu, Guoping Wang, and Sheng Li,
\newblock ``{{DiffSound}}: {{Differentiable}} modal sound rendering and inverse rendering for diverse inference tasks,''
\newblock in {\em {{ACM SIGGRAPH}} 2024 Conference Papers}, New York, NY, USA, 2024, Siggraph '24, Association for Computing Machinery.

\bibitem{diaz_rigidbody_2023}
Rodrigo Diaz, Ben Hayes, Charalampos Saitis, Gy{\"o}rgy Fazekas, and Mark Sandler,
\newblock ``Rigid-{{Body Sound Synthesis}} with {{Differentiable Modal Resonators}},''
\newblock in {\em {{ICASSP}} 2023 - 2023 {{IEEE International Conference}} on {{Acoustics}}, {{Speech}} and {{Signal Processing}} ({{ICASSP}})}, June 2023, pp. 1--5.

\bibitem{diaz_efficient_2024}
Rodrigo Diaz, Carlos De~La Vega~Martin, and Mark Sandler,
\newblock ``Towards {{Efficient Modelling}} of {{String Dynamics}}: {{A Comparison}} of {{State Space}} and {{Koopman}} based {{Deep Learning Methods}},''
\newblock in {\em Proc. {{Int}}. {{Conf}}. {{Digital Audio Effects}} ({{DAFx-24}})}, 2024.

\bibitem{lee_differentiable_2024}
Jin~Woo Lee, Jaehyun Park, Min~Jun Choi, and Kyogu Lee,
\newblock ``Differentiable {{Modal Synthesis}} for {{Physical Modeling}} of {{Planar String Sound}} and {{Motion Simulation}},''
\newblock in {\em The {{Thirty-eighth Annual Conference}} on {{Neural Information Processing Systems}}}, Nov. 2024.

\bibitem{trautmann_sound_2000}
L.~Trautmann and Rudolf Rabenstein,
\newblock ``Sound {{Synthesis}} with {{Tension Modulated Nonlinearities Based}} on {{Functional Transformations}},''
\newblock in {\em Proceedings of the {{Conference}} on {{Acoustics}} and {{Music}}: {{Theory}} and {{Applications}}}, Montego Bay, Jamaica, Dec. 2000.

\bibitem{hairer_geometric_2003}
Ernst Hairer, Christian Lubich, and Gerhard Wanner,
\newblock ``Geometric numerical integration illustrated by the {{St{\"o}rmer}}--{{Verlet}} method,''
\newblock {\em Acta Numerica}, vol. 12, pp. 399--450, May 2003.

\bibitem{bradbury_jax_2018}
James Bradbury, Roy Frostig, Peter Hawkins, Matthew~James Johnson, Chris Leary, Dougal Maclaurin, George Necula, Adam Paszke, Jake VanderPlas, Skye {Wanderman-Milne}, and Qiao Zhang,
\newblock ``{{JAX}}: Composable transformations of {{Python}}+{{NumPy}} programs,'' 2018,
\newblock Available at \href{http://github.com/jax-ml/jax}{http://github.com/jax-ml/jax}.

\bibitem{torres_unsupervised_2024}
Bernardo Torres, Geoffroy Peeters, and Ga{\"e}l Richard,
\newblock ``Unsupervised {{Harmonic Parameter Estimation Using Differentiable DSP}} and {{Spectral Optimal Transport}},''
\newblock in {\em {{ICASSP}} 2024 - 2024 {{IEEE International Conference}} on {{Acoustics}}, {{Speech}} and {{Signal Processing}} ({{ICASSP}})}, Apr. 2024, pp. 1176--1180.

\bibitem{blelloch_prefix_1990}
Guy~E Blelloch,
\newblock ``Prefix sums and their applications,''
\newblock Tech. {R}ep., School of Computer Science, Carnegie Mellon University, 1990.

\bibitem{kehling_idmtsmtguitar_2023}
Christian Kehling, Andreas M{\"a}nnchen, and Arndt Eppler,
\newblock ``{{IDMT-SMT-Guitar Dataset}},'' Jan. 2023,
\newblock Available at \href{https://zenodo.org/records/7544110}{https://zenodo.org/records/7544110}.

\bibitem{clarke_realimpact_2023}
Samuel Clarke, Ruohan Gao, Mason Wang, Mark Rau, Julia Xu, Jui-Hsien Wang, Doug~L. James, and Jiajun Wu,
\newblock ``{{RealImpact}}: A dataset of impact sound fields for real objects,''
\newblock in {\em Proceedings of the {{IEEE}} International Conference on Computer Vision and Pattern Recognition}, 2023.

\bibitem{turian_im_2020a}
Joseph Turian and Max Henry,
\newblock ``I'm {{Sorry}} for {{Your Loss}}: {{Spectrally-Based Audio Distances Are Bad}} at {{Pitch}},''
\newblock in {\em ''{{I Can}}'t {{Believe It}}'s {{Not Better}}!'' {{NeurIPS}} 2020 Workshop}, Dec. 2020.

\bibitem{dellasanta_automatic_2024}
Francesco Della~Santa,
\newblock ``Automatic {{Differentiation-Based Multi-Start}} for {{Gradient-Based Optimization Methods}},''
\newblock {\em Mathematics}, vol. 12, no. 8, pp. 1201, Jan. 2024.

\bibitem{ducceschi_nonlinear_2014}
Michele Ducceschi, Cyril Touz{\'e}, Stefan Bilbao, and Craig~J. Webb,
\newblock ``Nonlinear dynamics of rectangular plates: Investigation of modal interaction in free and forced vibrations,''
\newblock {\em Acta Mechanica}, vol. 225, no. 1, pp. 213--232, Jan. 2014.

\end{thebibliography}



\end{document}